\documentclass[reprint,superscriptaddress]{revtex4-1}
\usepackage{amsmath,graphicx,upgreek}
\renewcommand{\thefigure}{\arabic{figure} (color online)}

\begin{document}
\title{Temperature-Induced Inversion of the Spin-Photogalvanic Effect in WTe\textsubscript{2} and MoTe\textsubscript{2}}
\author{Sejoon Lim}
\affiliation{Department of Applied Physics, Stanford University, Stanford, California 94305, USA}
\affiliation{Stanford Institute for Materials and Energy Sciences, SLAC National Accelerator Laboratory,\protect\\2575 Sand Hill Road, Menlo Park, California 94025, USA}
\author{Catherine R. Rajamathi}
\affiliation{Max Planck Institute for Chemical Physics of Solids, 01187 Dresden, Germany}
\author{Vicky S{\"{u}\ss}}
\affiliation{Max Planck Institute for Chemical Physics of Solids, 01187 Dresden, Germany}
\author{Claudia Felser}
\affiliation{Max Planck Institute for Chemical Physics of Solids, 01187 Dresden, Germany}
\author{Aharon Kapitulnik}
\affiliation{Department of Applied Physics, Stanford University, Stanford, California 94305, USA}
\affiliation{Stanford Institute for Materials and Energy Sciences, SLAC National Accelerator Laboratory,\protect\\2575 Sand Hill Road, Menlo Park, California 94025, USA}
\affiliation{Department of Physics, Stanford University, Stanford, California 94305, USA}
\begin{abstract}
We investigate the generation and temperature-induced evolution of optically-driven spin photocurrents in WTe\textsubscript{2} and MoTe\textsubscript{2}. By correlating the scattering-plane dependence of the spin photocurrents with the symmetry analysis, we find that a sizeable spin photocurrent can be controllably driven along the chain direction by optically exciting the system in the high-symmetry $y$-$z$ plane. Temperature dependence measurements show that pronounced variations in the spin photocurrent emerge at temperatures that coincide with the onset of anomalies in their transport and optical properties. The decreasing trend in the temperature dependence starting below $\sim$150~K is attributed to the temperature-induced Lifshitz transition. The sign inversion of the spin photocurrent, observed around 50~K in WTe\textsubscript{2} and around 120~K in MoTe\textsubscript{2}, may have its origin in an interaction that involves multiple kinds of carriers.
\end{abstract}
\date{\today}
\maketitle

In recent years, the semimetallic transition-metal dichalcogenides, WTe\textsubscript{2} and MoTe\textsubscript{2}, have provided a platform for realizing unusual physical phenomena, including the nonsaturating extremely large magnetoresistance \cite{mAli14,dRhodes17}, the pressure-induced enhancement of superconductivity \cite{dKang15,xPan15,yQi16}, and the type-II Weyl semimetal states \cite{aSoluyanov15,fBruno16,yWu16,bFeng16,cWang16,ySun15,zWang16,lHuang16,kDeng16,jJiang17,aTamai16}.
Extensive research has been directed at uncovering the physical mechanisms that lead to these properties and their relation to the topological nature of the Weyl nodes. In the process, both compounds have been revealed to be governed by a complicated electronic structure that is characterized by an anomalous temperature dependence.

In the case of WTe\textsubscript{2}, for example, angle-resolved photoemission spectroscopy showed that compensation between electron and hole densities observed at low temperatures rapidly breaks down with increasing temperature \cite{iPletikosic14,yWu15}, and this shift of the Fermi level leads to a Lifshitz transition, suppressing the hole pockets above $\sim$160~K \cite{yWu15}. The magnetotransport study found a pronounced enhancement of both the effective mass anisotropy and the magnetoresistance below $\sim$75 K, which suggests yet another possible change in the electronic structure \cite{lThoutam15}. Hall and thermoelectric transport studies observed significant changes in the Hall coefficient below $\sim$50~K \cite{yLuo15,yWang15} and sign inversion of the thermoelectric power around 60~K \cite{sKabashima66}, respectively. Additional evidence for the change in the electronic structure around 50~K was provided by an ultrafast transient reflectivity measurement, which revealed an unusual decrease in the time scale of a relaxation process that was attributed to phonon-assisted electron-hole recombination \cite{yDai15}.

\begin{figure}
    \includegraphics[width=\linewidth]{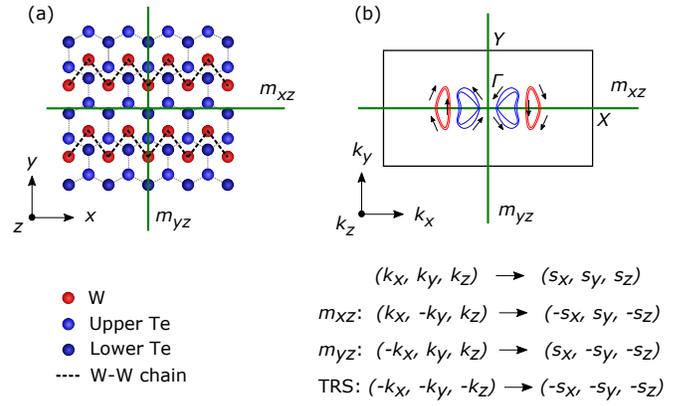}
    \caption{\label{fig:structure}(a) Top schematic view of the crystal structure of a single Te-W-Te compound layer in WTe\textsubscript{2} (space group $Pmn2_1$). The green lines, labeled $m_{xz}$ and $m_{yz}$, represent the glide reflection symmetry in the $x$-$z$ plane and the mirror symmetry in the $y$-$z$ plane, respectively. (b) Schematics of the Fermi surface and its spin texture in the $k_z=0$ plane of the Brillouin zone, adopted from previous spin- and angle-resolved photoemission studies \cite{jJiang15,bFeng16}. The red and blue lines indicate the electron and hole pockets, respectively. The black arrows indicate their spin texture, which necessarily respects $m_{xz}$, $m_{yz}$, and TRS, simultaneously. Effects of the symmetry operations on spins are summarized.}
\end{figure}

Further enriched with spin-splitting of bands due to strong spin-orbit coupling and broken inversion symmetry as illustrated in Fig.~1, spin-photogalvanics \cite{sGanichev03rev,eIvchenko17}, in which optical orientation with circularly polarized radiation leads to a spin-polarized electric current via the circular photogalvanic and the spin-galvanic effects, may provide insight into temperature-dependent dynamic properties of spin-polarized carriers in these compounds. Previously, this technique has been applied to study point group symmetries and dynamic properties of helicity-dependent carriers in semiconductors \cite{vBelinicher80}, quantum well structures \cite{sGanichev03rev}, topological insulators \cite{jMoore10,pHosur11,jMcIver12}, semiconducting transition-metal dichalcogenides \cite{hYuan14}, and recently type-I Weyl semimetals \cite{hIshizuka16,cChan17,fJuan17,qMa17}.

In this Letter, we report a detailed study of the generation and temperature-induced evolution of optically-driven spin photocurrents in WTe\textsubscript{2} and MoTe\textsubscript{2}. We first show, by varying the direction of exciting radiation, that the mirror symmetries along with time-reversal symmetry (TRS) forbid the generation of a spin photocurrent under certain experimental geometries. Its correlation with the symmetry analysis suggests that the generating mechanisms are rooted in the electronic structure and its spin texture, which necessarily respect the crystal symmetries. Subsequent temperature dependence measurements reveal sign inversion of the spin photocurrent around 50~K in WTe\textsubscript{2} and around 120~K in MoTe\textsubscript{2}. Along with the trend in the temperature dependence, which features a drop in the photocurrent magnitude starting below the Lifshitz transition temperature around 150~K, these observations suggest that the anomalous temperature dependence of the spin photocurrent in these compounds may be associated with the temperature-induced evolution of their electronic structure.

We first distinguish the photocurrent response that arises from the photogalvanic effects from the photoinduced thermoelectric response \cite{jMcIver12}. Figures 2(a) and 2(b) show the variation of the electric current, induced along the $x$ axis of WTe\textsubscript{2}, as the obliquely incident $p$-polarized light in the $y$-$z$ plane is scanned across the contacts at room temperature \cite{suppl}. The observation of the current flow away from the heated contacts, which is consistent with the $n$-type thermoelectric behavior of WTe\textsubscript{2} at room temperature \cite{sKabashima66}, along with the reversal of the current direction upon sweeping the light across the contacts, indicates the presence of a large thermoelectric contribution in the photoinduced electric current. As shown in the polarization dependence measurements in Fig.~2(c), however, this thermoelectric contribution is mostly independent of the polarization of incident radiation, which allows us to selectively investigate the photocurrent response by analyzing its polarization dependence. This analysis is further supported by the observation of the same $n$-type thermoelectric behavior, yet a sign-flipped spin photocurrent, upon rotating the $y$-$z$ scattering plane by 180$^\circ$ about the $z$ axis, as shown in Fig.~2(d).

\begin{figure}
    \includegraphics[width=\linewidth]{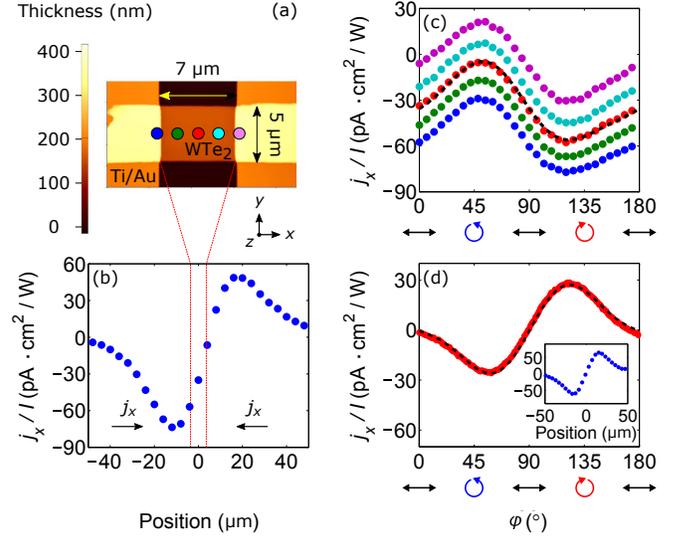}
    \caption{\label{fig:position}(a) Atomic force microscopy image of the measured 160~nm thick flake of WTe\textsubscript{2}. (b) Variation of the photoinduced electric current $j_x$, normalized by the light intensity $I$, as the obliquely incident $p$-polarized light in the $y$-$z$ plane is scanned across the contacts at room temperature. (c) Phase angle dependence of $j_x/I$. Each colored plot represents a measurement carried out with the incident radiation centered at the corresponding location shown in (a). The dashed black line represents a fit to the phenomenological expression. (d) Same measurements as (b) and (c), but with the $y$-$z$ scattering plane rotated by $180^\circ$ about the $z$ axis.}
\end{figure}

The observed polarization dependence can be described by a phenomenological expression, given by $j_\lambda = \sum_{\mu \nu} \chi_{\lambda \mu \nu} \left( E_\mu E_\nu^* + E_\nu E_\mu^* \right)/2 + \sum_{\mu} \gamma_{\lambda \mu} i \left( \boldsymbol{E} \times \boldsymbol{E}^* \right)_\mu + \sum_{\mu} Q_{\lambda \mu} S_{\mu} + \sum_{\delta \mu \nu} T_{\lambda \delta \mu \nu} q_\delta E_\mu E_\nu^*$, where $\boldsymbol{E}$ is the complex amplitude of the electric field of incident radiation, $\boldsymbol{q}$ is its wavevector inside the crystal, and $\boldsymbol{S}$ is the nonequilibrium average spin polarization \cite{sGanichev03rev}. The set of symmetry elements contained in the point group $C_{2v}$ of WTe\textsubscript{2} is encoded in the tensors, $\chi_{\lambda \mu \nu}$, $\gamma_{\lambda \mu}$, $Q_{\lambda \mu}$, and $T_{\lambda \delta \mu \nu}$, which describe the linear and circular photogalvanic, the spin-galvanic, and the photon drag effects, respectively. In terms of the phase angle $\varphi$ between the polarization axis of the laser radiation and the optical axis of the $\lambda/4$ plate, the photocurrent can be expressed as $j_x = j_C \sin 2\varphi + j_{L1} \sin 4\varphi + j_{L2} \cos 4\varphi + j_0$ with four scaling parameters, where the first term, which is proportional to the helicity $P_{circ} = \sin 2\varphi \propto \lvert \boldsymbol{E} \times \boldsymbol{E}^* \rvert \propto \lvert \boldsymbol{S} \rvert$, mainly arises from the circular photogalvanic and the spin-galvanic effects. In this study, we restrict our attention to these spin-related photogalvanic effects, which are parametrized by a single scaling parameter $j_C$.

\begin{figure*}
    \includegraphics[width=\linewidth]{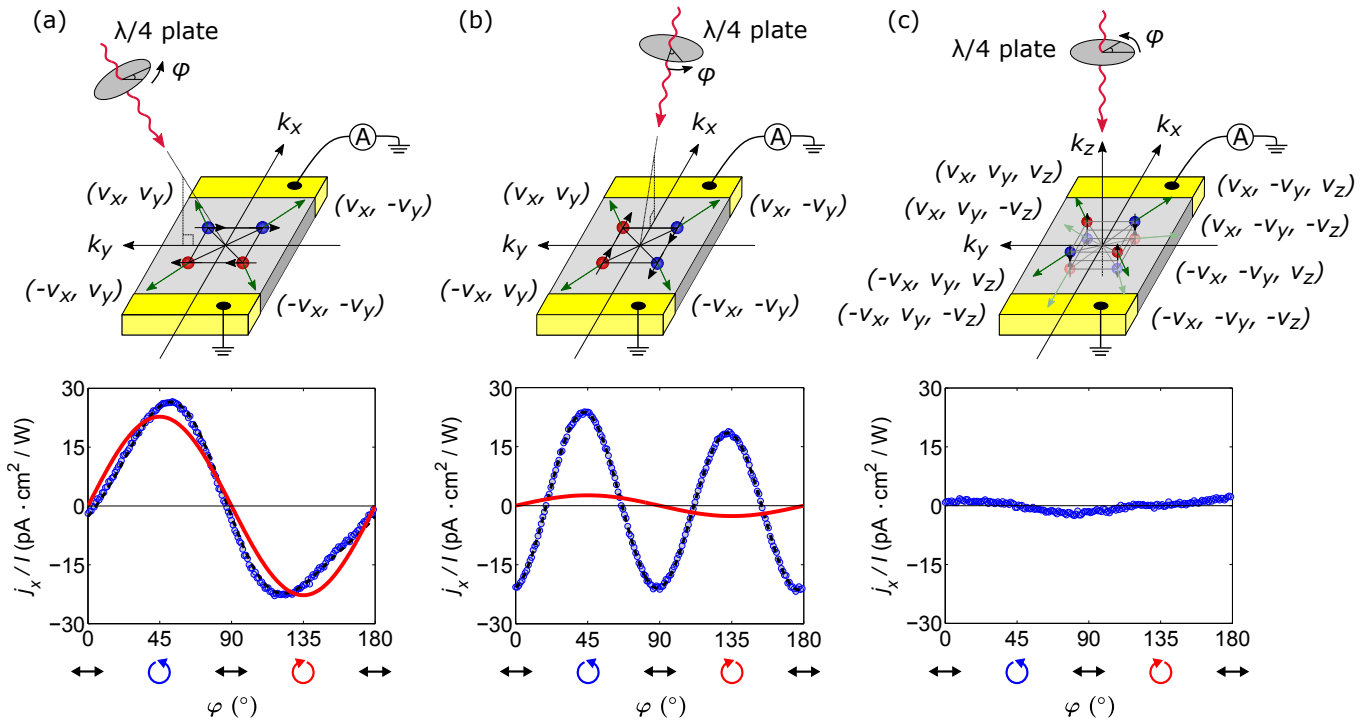}
    \caption{\label{fig:spin}(a) Sketch of the experimental geometry and the phase angle dependence of $j_x/I$ in WTe\textsubscript{2} with the light obliquely incident in the $y$-$z$ plane at an angle of 40$^\circ$. In the schematic, the four symmetry-related states in the $k_z = 0$ plane are shown with their spins marked with black arrows. The blue and red fillings indicate those that are preferentially excited with right- and left-handed helicity, respectively. Their group velocities are indicated by long green arrows and described on the sides. In the data figure, the open blue circles represent the photocurrent $j_x/I$ measured at room temperature with the polarization-independent component removed. The dashed black line shows a fit to the phenomenological expression. The solid red line is the extracted spin-dependent component, which is proportional to $P_{circ}$. (b) Same as (a), but with the light obliquely incident in the $x$-$z$ plane. (c) Same as (a), but with the light normally incident.}
\end{figure*}

Below, we investigate how the mirror and time-reversal symmetries of WTe\textsubscript{2} allow only the transverse spin photocurrent when the light is incident in the high-symmetry $x$-$z$ or $y$-$z$ plane using spin-selective optical excitation as the generating mechanism of spin photocurrents. Analogous data on MoTe\textsubscript{2} can be found in the Supplemental Material \cite{suppl}.

Figure~3 shows the photocurrent measured at room temperature under three different experimental geometries. When the light is incident in the $y$-$z$ plane, a large spin photocurrent develops along the $x$ axis, as shown in Fig.~3(a). In contrast, this spin photocurrent significantly diminishes, as shown in Fig.~3(b), when the scattering plane is rotated into the $x$-$z$ plane. On a microscopic level, these observations indicate that an asymmetric distribution of photoexcited carriers is induced predominantly about the high-symmetry scattering plane in $k$ space as a result of optical orientation \cite{sGanichev03rev,eIvchenko17}.

To understand this asymmetric photoexcitation, we note that each $k$ point in the Brillouin zone of WTe\textsubscript{2} is related to seven other points by the mirror and time-reversal symmetries, whose constraints on the spin texture are described in Fig.~1(b). For simplicity, we consider here the $k$ points lying in the $k_z = 0$ plane of the Brillouin zone, which have only in-plane spin components and four symmetry-related states. As illustrated in the sketches in Figs.~3(a) and 3(b), depending on its helicity, the incident circularly polarized radiation excites the occupied states asymmetrically about the scattering plane. In contrast, photoexcitation is symmetric about the other high-symmetry plane, which results in a net spin-polarized photocurrent in the direction normal to the scattering plane. This analysis, based on crystal symmetries, can also account for the absence of a spin photocurrent under normal incidence, which is described in Fig.~3(c). It is also applicable when $k$-dependent spin-flip scattering of photoexcited carriers, instead of spin-selective photoexcitation, is considered as the generating mechanism of spin photocurrents.

Having identified the electronic structure and its spin texture as the bases for determining the properties of spin photocurrents, we proceed to investigate how the spin photocurrent in WTe\textsubscript{2} evolves as its electronic structure undergoes transitions with temperature. Figure~4 shows this temperature-induced evolution, quantified by the scaling parameter $j_C$. As shown in the lower inset, our light intensity of $\sim$42~W/cm$^2$ lies in the range where the spin photocurrent scales linearly with the intensity, which signals that contributions from higher-order nonlinear effects are small. Above $\sim$150~K, the spin photocurrent stays relatively constant, which likely indicates that the transition probabilities and the momentum and spin relaxation times are not strongly affected by the Fermi level shift and thermal broadening in this temperature range. As temperature is lowered below $\sim$150~K, the spin photocurrent starts to decrease and this decreasing behavior accelerates. Intriguingly, this onset roughly coincides with the Lifshitz transition around 160~K, below which the hole bands begin to surface above the Fermi level, as schematically illustrated in the upper inset of Fig.~4. This coincidence suggests that the hole pockets may play a role in generating and scattering the photoexcited carriers.

\begin{figure}
    \includegraphics[width=\linewidth]{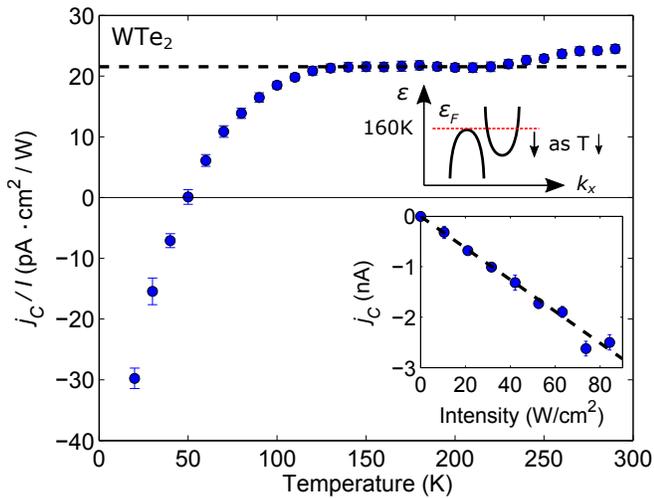}
    \caption{\label{fig:wte2temperature}Temperature dependence of the scaling parameter $j_C$ in WTe\textsubscript{2}. The horizontal dashed black line has been added as a guide to the eyes. The lower inset shows its intensity dependence at 20~K. The dashed black line in the inset represents a linear fit. The upper inset shows a schematic of the band structure along $\Gamma$-$X$. Around 160~K, the Fermi level, represented by the dashed red line, crosses the top of the hole band.}
\end{figure}

One possible mechanism is through an enhancement of interband relaxation of the photoexcited carriers. As the number of holes increases with the appearance of the hole pockets in the Fermi surface, both the radiative and nonradiative phonon-assisted recombination processes become more efficient, which facilitates interband cooling \cite{yDai15}. The resulting drive to restore the nonequilibrium carrier distribution back to equilibrium would then tend to increase the probability rates for optical transitions and decrease the momentum and spin relaxation times, thereby influencing both the circular photogalvanic and the spin-galvanic processes \cite{sGanichev03rev,eIvchenko17}.

Upon lowering the temperature further, the sign of the scaling parameter $j_C$ flips around 50~K, which indicates that the net spin photocurrent reverses its direction of flow. Such inversion at fixed helicity has frequently been observed in spectral responses, where variations of the photon energy shift optical transitions across band minima \cite{sGanichev03,lGolub03,cYang06}. In contrast, temperature-induced sign inversion has been rarely demonstrated \cite{sGanichev00}. Its observation in $p$-GaAs/AlGaAs quantum wells \cite{sGanichev00} has been attributed to a change of the scattering mechanism from phonons to impurities; however, it may have a different origin in the case of WTe\textsubscript{2}, because the onset temperature is high. Based on the aforementioned transport and optical studies pointing toward a possible change in the electronic structure around 50~K and a plausible explanation for the temperature dependence of the Hall coefficient and the thermoelectric power given by a three-carrier model \cite{sKabashima66}, it may be likely that this inversion arises from an interaction that involves multiple kinds of carriers.

As shown in Fig.~5, similar sign inversion is observed in MoTe\textsubscript{2} around 120~K, which again coincides with the temperature below which its thermoelectric power crosses over to a $p$-type behavior \cite{fChen16,dRhodes17}. In contrast to WTe\textsubscript{2}, however, this temperature range is higher than the onset temperature, around 70~K, of other anomalies present in MoTe\textsubscript{2}, such as in the effective mass anisotropy \cite{fChen16}, the magnetoresistance \cite{fChen16}, the Hall coefficient \cite{qZhou16,fChen16,dRhodes17}, and the specific heat \cite{dRhodes17}, which suggests that the multiple-carrier picture may have the dominant effect. The decreasing trend of the scaling parameter $j_C$ below $\sim$150~K may again be associated with a possible Lifshitz transition, which was suggested by previous Hall and thermoelectric transport studies \cite{fChen16}, although the onset is not as sharp as in WTe\textsubscript{2} possibly due to the substantial thermal hysteresis associated with the structural transition, which extends from $\sim$125~K to room temperature \cite{rClarke78,dRhodes17}.

\begin{figure}
    \includegraphics[width=\linewidth]{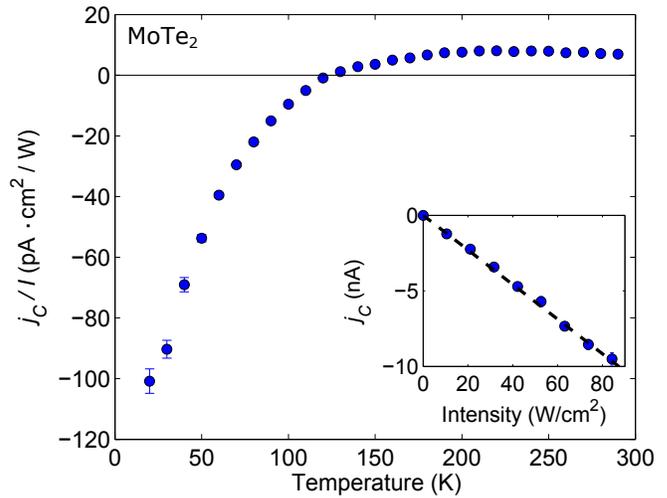}
    \caption{\label{fig:mote2temperature}Temperature dependence of the scaling parameter $j_C$ in MoTe\textsubscript{2}. The inset shows its linear intensity dependence at 20~K.}
\end{figure}

We also note that the spin photocurrent is small but nonzero above $\sim$250~K, where MoTe\textsubscript{2} is known to crystallize in the centrosymmetric $1T'$ phase \cite{rClarke78,ySun15,hSakai16}. Although the absence of strong Raman modes that correspond to the semiconducting $2H$ phase or the noncentrosymmetric $T_d$ phase in the polarized Raman spectrum \cite{suppl} seems to rule out the possibility of structural transitions induced by strain \cite{ySun15,zWang16}, doping \cite{hSakai16,dRhodes17sub}, or pressure \cite{yQi16,hTakahashi17}, it may be possible that the regions near the evaporated contacts experience more deformation. We also suspect that there may be some contributions from the surface layers, which necessarily break inversion symmetry.

In summary, we find pronounced temperature-induced variations of the optically-driven spin photocurrents in WTe\textsubscript{2} and MoTe\textsubscript{2}, whose origin may be associated with anomalies in their transport and optical properties. In particular, we showed that sign inversion at fixed helicity can be induced by varying the temperature. Its rough overlap in temperature with the $n$-type to $p$-type crossover in the thermoelectric power in both compounds supports microscopic mechanisms based on a multiple-carrier picture. However, due to the numerous other anomalies that emerge at nearby temperatures, a detailed understanding of their correlation may be crucial in elucidating the exact origin of the sign inversion.

This work was supported by the Department of Energy, Office of Science, Basic Energy Sciences, Materials Sciences and Engineering Division, under Contract DE-AC02-76SF00515. Part of this work was performed at the Stanford Nano Shared Facilities with support from the Gordon and Betty Moore Foundation's EPiQS Initiative through Grant GBMF4529.

\newpage
\widetext

\begin{center}
\textbf{\large Supplemental Material:\\Temperature-Induced Inversion of the Spin-Photogalvanic Effect in WTe\textsubscript{2} and MoTe\textsubscript{2}}
\end{center}
\setcounter{figure}{0}
\renewcommand{\thefigure}{S\arabic{figure} (color online)}
\makeatletter

\section{Crystal growth}
Polycrystalline WTe\textsubscript{2} was synthesized by annealing finely ground stoichiometric amounts of W and Te elements (total weight of 4~g) in an evacuated quartz tube at 800~$^\circ$C at a rate of 30~$^\circ$C/hour for 7~days, followed by cooling slowly to the ambient temperature. Single crystals of WTe\textsubscript{2} were then grown by chemical vapour transport of the polycrystalline WTe\textsubscript{2} powder using TeCl\textsubscript{4} as a transport agent. 1~g of this polycrystalline powder and TeCl\textsubscript{4} (3~mg~ml$^{-1}$) were sealed in a quartz ampoule, which was then flushed with Ar, evacuated, sealed and heated in a two-zone furnace. Crystallization was conducted from (T\textsubscript{2}) 930~$^\circ$C to (T\textsubscript{1}) 880~$^\circ$C.

A similar chemical vapour transport method was employed to grow MoTe\textsubscript{2} crystals using the polycrystalline MoTe\textsubscript{2} powder and TeCl\textsubscript{4} as a transport additive. Polycrystalline MoTe\textsubscript{2} was synthesized by annealing stoichiometric amounts of Mo and Te elements at 800~$^\circ$C. The following crystallization was conducted from (T\textsubscript{2}) 1,000~$^\circ$C to (T\textsubscript{1}) 900~$^\circ$C. The quartz ampoule was then quenched in ice water to yield the high-temperature monoclinic phase, the $1T'$-MoTe\textsubscript{2}. When cooling down below around 250~K, a structural phase transition occurs from the centrosymmetric $1T'$ phase to the noncentrosymmetric $T_d$ phase, which is the Weyl semimetal phase.

\section{Experimental setup}
The spin photocurrents were measured along the $x$ crystallographic axis of WTe\textsubscript{2} using two-terminal devices fabricated out of flakes mechanically exfoliated from a bulk single crystal onto a silicon wafer with 300 nm of thermal oxide. We worked with flakes thicker than the optical penetration depth of $\sim$110~nm \cite{aFrenzel17,cHomes15} in order to minimize contributions from the radiation reflected from the substrate, which necessarily has flipped helicity. The two-terminal system without an external bias was homogeneously irradiated with 1550~nm (0.80~eV) continuous wave light, obliquely incident in the $y$-$z$ plane at an angle of $40^\circ$, as schematically illustrated in Fig.~S2(a). The incident $p$-polarized radiation from the laser diode was focused onto a $\sim$68 $\upmu$m spot on the sample to yield $\sim$42 W/cm$^2$ of intensity, and its polarization was modified to circular polarization by applying a $\lambda/4$ plate. The degree of circular polarization varies according to $P_{circ} = \sin 2\varphi$, where $\varphi$ is the phase angle between the polarization axis of the laser radiation and the optical axis of the $\lambda/4$ plate, with $P_{circ} = \pm 1$ corresponding to right- and left-handed helicity, respectively \cite{sGanichev03rev}. In order to account for slight deflections of the incident light caused by the rotating $\lambda/4$ plate, we averaged the data from $\varphi = 0^\circ \sim 180^\circ$ with $\varphi = 180^\circ \sim 360^\circ$. We note that the variations in the beam size and location can somewhat change the magnitude of the spin photocurrents, but not the trend in its temperature dependence, which governs the physics that we study here.

\section{Determination of the crystal axes}
Polarized Raman spectroscopy was used to identify the crystal axes of WTe\textsubscript{2} and MoTe\textsubscript{2} flakes. 638 nm excitation light was focused onto a $\sim$3 $\upmu$m spot on the sample, and its Raman response was measured in the parallel-polarization configuration, where the polarizations of the incident and scattered light are parallel to each other. The angular dependence of the mode intensities was obtained by rotating the flakes with the light polarizations fixed.

Figure S1(a) shows the Raman spectrum of the WTe\textsubscript{2} flake that was measured in the photogalvanics experiment. The measurement was carried out at room temperature with the flake oriented at $\theta \sim 45^\circ$. The angle $\theta$, which represents counterclockwise rotation of the flake, is defined in such a way that, at $\theta = 0^\circ$, a line joining the electrical contacts is parallel to the light polarization axis. The observed Raman modes are in good agreement with previous Raman scattering studies \cite{qSong16}. In the parallel-polarization configuration, the intensity of the mode at $\sim$164 cm$^{-1}$ ($\sim$212 cm$^{-1}$) has been reported to exhibit a two-lobe shaped angular dependence in its polar plot and to reach its maximum (minimum) value when the \mbox{W-W} chains are aligned parallel to the light polarization axis. As demonstrated in Fig.~S1(b), we analyzed the intensity ratio of these two modes, $I_{164\,cm^{-1}}/I_{212\,cm^{-1}}$, to determine the crystal axes. It is evident that the contacts lie along the $x$ axis of WTe\textsubscript{2}.

A similar analysis was carried out on the MoTe\textsubscript{2} flake at room temperature, as shown in Figs.~S1(c) and S1(d), in which case the Raman mode at $\sim$163 cm$^{-1}$ ($\sim$78 cm$^{-1}$) is known to reach its maximum (minimum) intensity when the \mbox{Mo-Mo} chains are aligned parallel to the light polarization axis \cite{qSong17,xMa16}. A comparison with previous Raman scattering studies on the $2H$ phase \cite{qSong17} and the $1T'$ phase \cite{xMa16} shows that there are no strong Raman modes that correspond to these phases in our spectrum.

\begin{figure}
    \includegraphics[width=\linewidth]{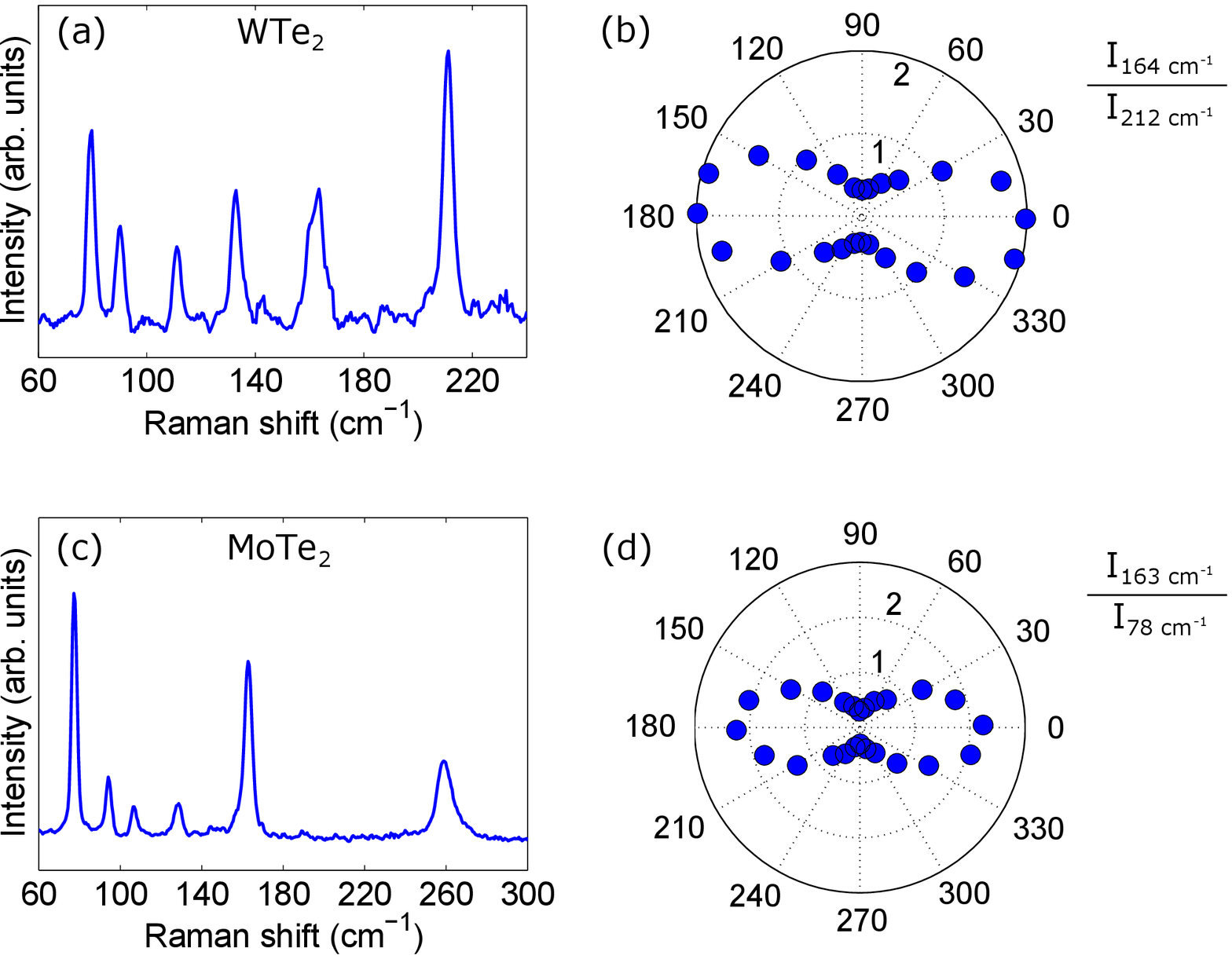}
    \caption{\label{fig:raman}(a) Room-temperature Raman spectrum of the WTe\textsubscript{2} flake measured in the parallel-polarization configuration at $\theta \sim 45^\circ$. (b) Angular dependence of the intensity ratio of the modes at $\sim$164 cm$^{-1}$ and $\sim$212 cm$^{-1}$. (c,d) Same measurements for the MoTe\textsubscript{2} flake.}
\end{figure}

\section{Scattering-plane dependence the spin photocurrent in \textit{T\textsubscript{\MakeLowercase{d}}}-M\MakeLowercase{o}T\MakeLowercase{e}\textsubscript{2}}
The low-temperature orthorhombic phase of MoTe\textsubscript{2}, known as the $T_d$-MoTe\textsubscript{2}, belongs to the same space group $Pmn2_1$ as WTe\textsubscript{2} and is necessarily characterized by the same symmetry properties \cite{ySun15}. Figure~S2 shows the photocurrent measured on MoTe\textsubscript{2} under three different experimental geometries at 20 K, analogous to Fig.~3 of the main text for WTe\textsubscript{2}. As expected from the same symmetry analysis, the spin photocurrent is generated predominantly in the direction normal to the high-symmetry scattering plane.

\begin{figure}
    \includegraphics[width=\linewidth]{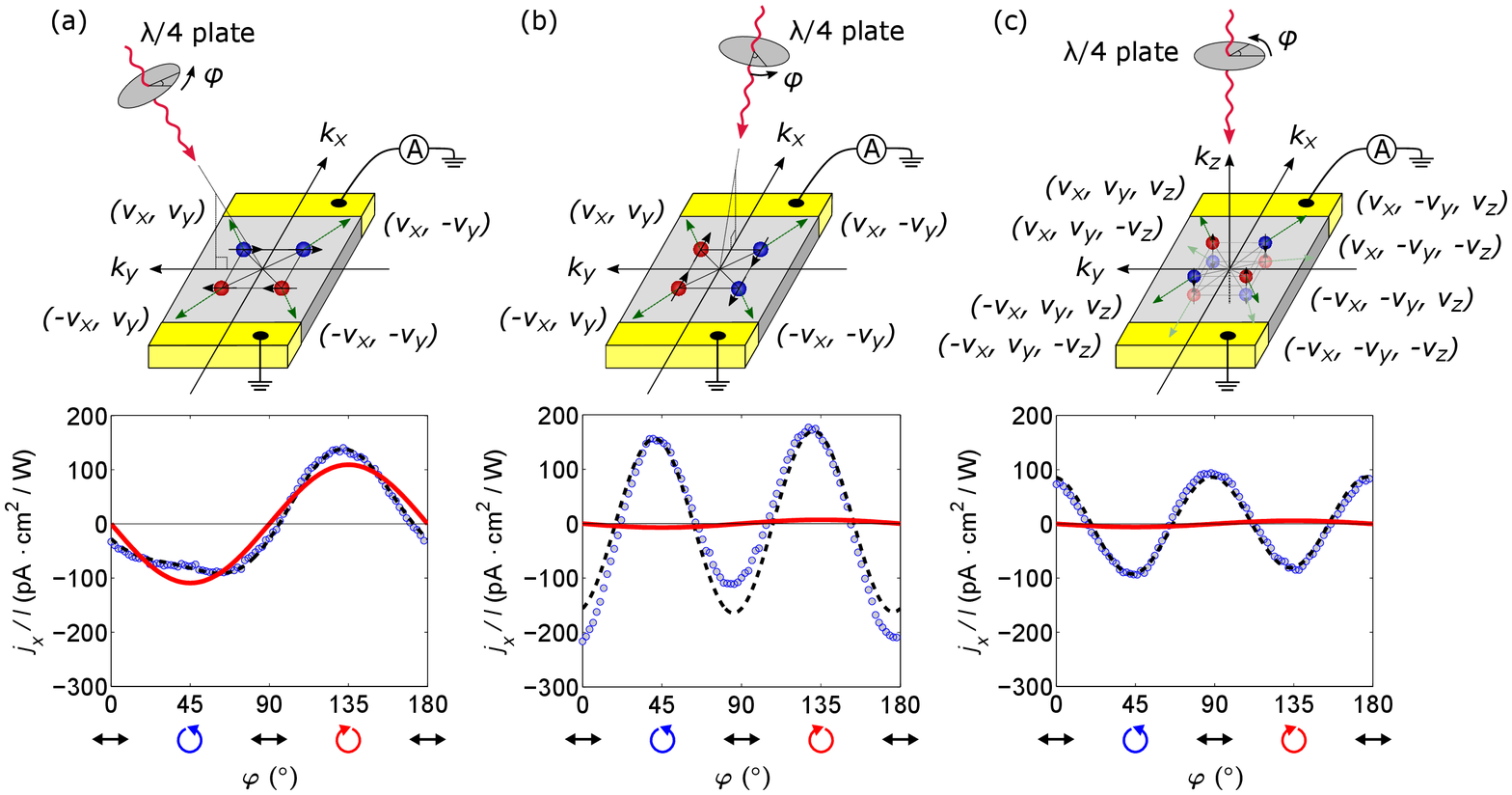}
    \caption{\label{fig:mote2spin}Polarization dependence measurements carried out on MoTe\textsubscript{2} at 20 K.}
\end{figure}

\end{document}